\documentclass[onecolumn,aps,prb,floatfix,10pt,amsmath,amssymb,notitlepage,longbibliography,superscriptaddress]{revtex4-2}
\usepackage{graphicx}
\usepackage{booktabs}
\usepackage{hyperref}
\pdfoutput=1
\newcommand{\tref} [1]{Table~\ref{#1}}
\newcommand{\fref} [1]{Fig.~\ref{#1}}
\newcommand{\Fref} [1]{Figure~\ref{#1}}

\newcommand{\sref} [1]{Sec.~\ref{#1}}

\newcommand{\eref} [1]{Eq.~\eqref{#1}}

\newcommand{\eeref}[1]{Eqs.~\eqref{#1}}
\newcommand{\EEref}[1]{Equations~\eqref{#1}}
\newcommand{\cref} [1]{ref.~[\onlinecite{#1}]}
\newcommand{\Cref} [1]{Reference~[\onlinecite{#1}]}

\begin{document}

\title{Regimes and quantum bounds of nanoscale thermoelectrics with peaked transmission function}
\author{Giuseppe Bevilacqua}
\affiliation{DIISM, Universit\`{a} di Siena, Via Roma 56, I-53100 Siena, Italy}
\author{Alessandro Cresti}
\affiliation{Univ. Grenoble Alpes, Univ. Savoie Mont Blanc, CNRS, Grenoble INP, IMEP-LAHC, 38000 Grenoble, France}
\author{Giuseppe Grosso}
\affiliation{Dipartimento di Fisica ``E. Fermi'', Universit\`{a} di Pisa, Largo Pontecorvo 3, I-56127 Pisa, Italy}
\author{Guido Menichetti}
\affiliation{Dipartimento di Fisica ``E. Fermi'', Universit\`{a} di Pisa, Largo Pontecorvo 3, I-56127 Pisa, Italy}
\affiliation{Istituto Italiano di Tecnologia, Graphene Labs, Via Morego, 30, I-16163 Genova, Italy}
\author{Giuseppe \surname{Pastori Parravicini}}
\affiliation{Dipartimento di Fisica ``E. Fermi'', Universit\`{a} di Pisa, Largo Pontecorvo 3, I-56127 Pisa, Italy}
\affiliation{Dipartimento di Fisica ``A. Volta'', Universit\`{a} di Pavia, Via A. Bassi, I-27100 Pisa, Italy}
%

\begin{abstract}
Based on the Landauer-B\"{u}ttiker theory, we explore the thermal regimes of two-terminal nanoscale systems with an energy-peaked transmission function. 
The device is in contact with two reservoirs held at different temperatures and chemical potentials. 
We identify the operation regions where the system acts as energy pump (thermal machine) or heat pump (refrigerator machine), or where it is working in dissipative modes. The corresponding thermoelectric parameters are obtained without numerical calculations. 
The recent literature, by focusing on systems with box-like or step- like shapes of the transmission functions, demonstrated that bounds of quantum origin exist for output power and heat currents of thermal machines and refrigerators. 
The simple model we adopt in this paper allows us to grasp easily and without numerical calculations the presence of quantum bounds for the above thermoelectric quantities, as function of the position of the transmission peak with respect to the chemical potentials of the left and right reservoirs. 
In spite of the simple model and treatment, our results are in qualitative agreement with analytic findings in previous researches obtained with more realistic description of the electronic transmission function. 
\end{abstract}

\maketitle


\section{Introduction\label{sec1}}
 
The development of nanotechnology has paved the way for new strategies to increase the efficiency of thermoelectric processes~\cite{Whitney2018}. 
The pioneering papers by Hicks and Dresselhaus~\cite{Hicks1993,Hicks1993a,Dresselhaus2007} evidenced the importance of investigating nanoscale quantum transport to enhance the thermoelectric dimensionless figure of merit $ZT$ in the linear regime. 
$ZT$ is defined as $ZT=\sigma S^2 T/(\kappa_{\rm el}+\kappa_{\rm ph})$, where $\sigma$ is the electronic conductance, $S$ the Seebeck coefficient, $T$ the absolute temperature, and $\kappa_{\rm el}$ ($\kappa_{\rm ph}$) the electronic (phononic) thermal conductance. 
Several strategies were reported to maximize $ZT$ by a suitable choice of device design and appropriate materials~\cite{Dmitriev2010,Zebarjadi2012,Neophytou2015,Zlatic2014,Mahan2016,Snyder2008,Dettori2015,Dunham2016,Culebras2014,Masood2018,Urban2019}. 
Most attempts proposed to increase phonon scattering so to decrease the lattice thermal conductivity by engineering nanostructured devices. Other attempts proposed to increase the power factor, $\sigma S^2$, by varying the concentration of charge carriers~\cite{Dmitriev2010}.

As alternative approach Mahan and Sofo~\cite{Mahan1996} addressed the problem in a formal way, looking for the material with an optimal transport coefficient $\mathcal T(E)$ that guarantees, at given lattice thermal conductivity, the highest $ZT$. A transmission coefficient with $\delta$-like shape turned out to be the ideal choice. 
Along this line, the impact of energy spectrum width~\cite{Humphrey2005,Luo2013} and of other $\mathcal T(E)$ shapes as box-, Lorentzian\cite{Breit1936} and Fano\cite{Miroshnichenko2010,Grosso2014} like features, were successively considered~\cite{Bevilacqua2016} depending on specific problems or suggested by quantum broadening effects due the contacts. In particular, sharp features in $\mathcal T(E)$ approaching $\delta$-shape were realized and analyzed in terms of narrow rectangular shapes\cite{Yu2016} or single Lorentzian peaks of vanishing width $\Gamma$~\cite{Liu2018,Luo2016}, in quantum dots weakly interacting with the contacts~\cite{Sothmann2015,Talbo2017,Menichetti2018} and in the presence of electron-electron interaction\cite{Zubov2018}, in single molecule junctions~\cite{Torres2015}, molecular electronics~\cite{Reddy2007,Lambert2016,Zhan2011}, and resonant tunneling devices~\cite{Patil2017}.

The weaknesses, for practical realizations, of the strictly $\delta$-shape transmission coefficient proposed by Mahan and Sofo~\cite{Mahan1996}, and its conceptual limit, have been fully evidenced in the literature~\cite{ Whitney2014,Whitney2015,Luo2018}. 
As shown by Whitney\cite{Whitney2014,Whitney2015}, in the absence of phonon contribution to thermal conductance, such a limitation can be overcome by maximizing the efficiency for given output power.
Most importantly he provided analytic expressions for quantum bounds of output power and heat currents for thermoelectric systems with box-like or step-like transmission functions.

In this paper we study the effects of a peaked transmission function on the thermoelectric transport properties of a nanostructured system, in the absence of lattice contribution to the thermal conductivity, and beyond the linear response regime. 
The nonlinear regime is commonly reached in low-dimensional systems where large values of temperature and electrical potential gradients may easily occur due to their size~\cite{Talbo2017,Yamamoto2015,Benenti2017,Azema2014,Whitney2013,Whitney2015,Hershfield2013,Sanchez2016,Jiang2016}. 

In the framework of the Landauer-B\"{u}ttiker theory, we show in \sref{sec2} that a system with an extremely sharp peak resonance at a given energy $E_d$ can behave as an efficient thermal machine or a refrigerator, or a (useless) energy sink, depending on temperatures and chemical potentials of the reservoirs. 
 From the expressions of exchanged power and heat currents we show in \sref{sec3} that it is possible to easily infer for them the existence of bounds of quantum origin. 
 Our results provide intuitive, transparent and analytic expressions for the existence of quantum bounds in the value of exchanged power and heat currents in agreement with general findings obtained in the literature through more laborious calculation adopting box-like or step-like shapes of the transmission coefficient~\cite{Whitney2014,Whitney2015}.
\sref{sec4} contains our conclusive remarks.
 
 
\section{Model details and thermoelectric transport equations in the non-linear response regime\label{sec2}}
   
 In this section, by the scattering formalism, we study the different thermoelectric regimes of transport through a two-terminal mesoscopic electronic system characterized by a peaked sharp transmission coefficient ${\mathcal T}(E)$, where $E$ is the electron energy.
 Lorentzian resonances and antiresonances, and Fano transmission profiles, which mimic more realistic situations including quantum interference effects and level broadening due to coupling with reservoirs~\cite{Ferry2009,Datta1997,Datta2005,Dubi2011} are not treated here. 

We assume without loss of generality, that the temperature of the left reservoir is hotter than the one of the reservoir, namely $T_{\rm L} > T_{\rm R}$. 
No a priori assumption is done on the chemical potentials $\mu_{\rm L} $ and $\mu_{\rm R}$ of the particle reservoirs.
 
According to the Landauer-B\"{u}ttiker theory, in steady state conditions, the $left$ or $right$ particle number current $I_N^{\rm (L,R)}$, charge (electric) current $I_e^{\rm (L,R)}$, and heat (thermal) currents $I_Q^{\rm (L,R)}$, are given respectively by the expressions:
\begin{subequations}  \label{eq01}
\begin{eqnarray}      
  I_N & = & I_N^{\rm (L)} \ = \ I_N^{\rm (R)}
	     \ = \  \frac{1}{h} \int_{-\infty}^{+\infty} dE \, {\mathcal T} (E) \left[ f_{\rm L}(E) - f_{\rm R}(E) \right] \label{eq01a} 
       \\ [2mm]
  I_e & = & I_e^{\rm (L)} \ = \ I_e^{\rm (R)} = -eI_N  \label{eq01b}
        \\ [2mm]
  I_{Q}^{\rm (L,R)} & = & \frac{1}{h} \int_{-\infty}^{+\infty} dE (E- \mu_{\rm L,R}) \, {\mathcal T}(E) \left[ f_{\rm L}(E) - f_{\rm R}(E) \right]  \label{eq01c} 
\end{eqnarray} 
where ($-e$) is the electron charge and $h$ the Planck constant.
Due to particle charge conservation, the left and right number currents are equal and the same holds for the left and right charge currents. On the contrary, the left and right heat currents can have different values. 
We adopt the choice of positive direction for the currents flowing from the left reservoir to the central device, and for those flowing from the central device to the right reservoir. 
The exchanged power ${\mathcal P}$ is given by
\begin{eqnarray}              
    {\mathcal P} &= & I_{Q}^{\rm (L)} - I_{Q}^{\rm (R)} \ = \ (\mu_{\rm R} - \mu_{\rm L}) I_N \ .  \label{eq01d} 
\end{eqnarray} 
\end{subequations} 
\EEref{eq01} are general, and apply both in the linear regime (small difference of chemical potentials and temperatures of the two reservoirs), and in the non-linear regime (arbitrary difference of the thermodynamic parameters of the two reservoirs).   
We assume that the relation between the applied bias potential and the reservoir chemical potentials is given by $(-e)(V_{\rm L}-V_{\rm R}) = (-e) \Delta V = \Delta \mu = \mu_{\rm L} -\mu_{\rm R}$. 
 
\begin{figure}[t] 
 \begin{center}
 \includegraphics[width=8cm]{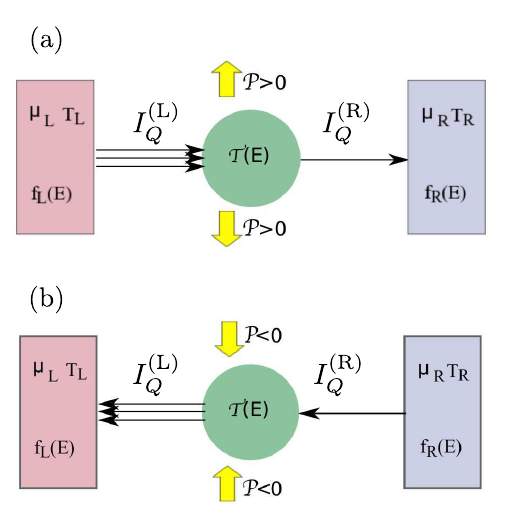}
 \end{center}
 \caption{\label{fig1} (a) Schematic representation of the two-terminal thermoelectric device in the power generation mode. 
 Heat extracted from the hot reservoir $(T_{\rm L}>T_{\rm R})$ is partially transferred to the cold reservoir, and the rest converted into usable power.
 (b) Schematic representation of a thermoelectric device in the refrigeration mode. Heat is extracted from the cold reservoir $(T_{\rm R}<T_{\rm L})$ and pumped into the hot reservoir with the absorption of external energy. } 
 \end{figure}

The \emph{power production mode} is characterized by the fact that \emph{left thermal current, right thermal current and available power, are all positive}. 
In this mode heat flows from the hot reservoir to the cold one, and part of the thermal energy is converted into available power, as schematically shown in \fref{fig1}(a). 
The efficiency of the device in the thermal machine mode is defined as
\begin{equation} \label{eq02}
  \eta^{(tm)} \ = \ \frac{ {\mathcal P} } { I_{Q}^{\rm (L)} }
              \ = \ \frac{ I_{Q}^{\rm (L)} - I_{Q}^{\rm (R)} } { I_{Q}^{\rm (L)} }
							\ \le \ \frac{ T_{\rm L} - T_{\rm R} } { T_{\rm L}}
							\ \equiv \  \eta_c^{(tm)} \ ,
\end{equation} 
where $ \eta_c^{(tm)}$ indicates the Carnot thermal efficiency.
The thermodynamic bounds of the thermal machine efficiency range from zero (for $T_{\rm L} \approx T_{\rm R}$) to unity (for $T_{\rm R} \ll T_{\rm L}$). 
 
The \emph{refrigeration mode} of the system is characterized by the fact that \emph{left thermal current, right thermal current and exchanged power are all negative}.
The heat is extracted from the cold reservoir $(T_{\rm R}<T_{\rm L})$ and pumped into the hot reservoir, with the absorption of external energy as schematically shown in \fref{fig1}(b).
In the refrigeration mode the machine efficiency is defined as 
\begin{equation} \label{eq03}
   \eta^{(refr)} \ = \ \frac{ I_{Q}^{\rm (R)} }{ \mathcal P }
                 \ = \  \frac{ I_{Q}^{\rm (L)} }{ I_{Q}^{\rm (L)} - I_{Q}^{\rm (R)} } - 1
								 \ \le \  \frac{T_{\rm L}}{T_{\rm L}-T_{\rm R}} -1
								 \ = \  \frac{T_{\rm R}}{T_{\rm L}-T_{\rm R}} 
   \equiv \eta_c^{(refr)} \ ,
\end{equation} 
which is clearly unbounded and diverges for $T_{\rm L}\approx T_{\rm R}$.

The efficiency indicated in \eref{eq02} and \eref{eq03} refers exclusively to thermal and refrigeration machines.
Neither $\eta^{(tm)}$ nor $\eta^{(refr)}$ has a clear physical meaning when the system is working in dissipative modes.
 
\begin{figure}[t]
 \begin{center}
 \includegraphics[width=8cm]{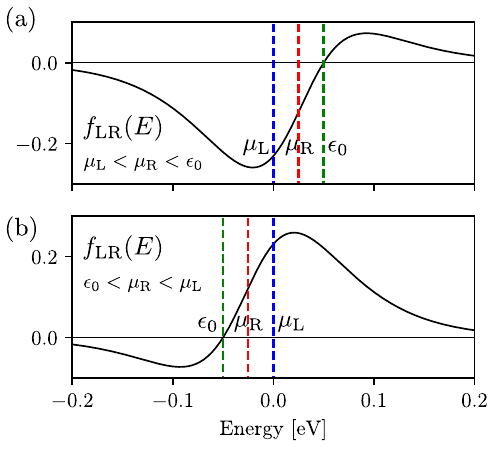}
 \end{center}
 \caption{ \label{fig2} Schematic representation of the Fermi function difference $ \, f_{\rm LR}(E) \equiv f_{\rm L}(E) - f_{\rm R}(E)$, for $T_{\rm L}=600$~K, $T_{\rm R}=300$~K. 
	(a) In the case $\mu_{\rm L}=0$~eV and $\mu_{\rm R}=0.025$~eV the sequence of variables on the energy axis is $\mu_{\rm L}<\mu_{\rm R} <\varepsilon_0= 0.05$~eV, $\varepsilon_0$ being the energy value for which $f_{\rm LR}(E) = 0$. 
	(b) In the case $\mu_{\rm L}=0$~eV and $\mu_{\rm R}=-0.025$~eV, the sequence of variables on the energy axis is $-0.05$~eV$=\varepsilon_0< \mu_{\rm R}<\mu_{\rm L}$.
 The vertical red, blue and green dashed lines indicate the position of $\mu_{\rm R} $, $\mu_{\rm L}$ and $\varepsilon_0$, respectively. } 
\end{figure}

From the transport \eeref{eq01}, it is apparent the central role played by ${\mathcal T}(E)$ and by the difference of the Fermi functions $f_{\rm LR}(E)\equiv f_{\rm L}(E)-f_{\rm R}(E)$, which is positive if
\begin{equation} \label{eq04}
 E \ > \ \varepsilon_0 \ \equiv \ \frac{\mu_{\rm R} \,T_{\rm L} - \mu_{\rm L} \,T_{\rm R}}{T_{\rm L} - T_{\rm R}}
   \ = \ \mu_{\rm R}+ \frac{T_{\rm R}}{T_{\rm L} - T_{\rm R}}(\mu_{\rm R}-\mu_{\rm L}) \ .
\end{equation} 
The function $f_{\rm LR}(E)$ has a single zero at $E=\varepsilon_0$, which is above both chemical potentials for $\mu_{\rm L}< \mu_{\rm R}$, and below both chemical potentials if $\mu_{\rm L} > \mu_{\rm R}$, see \fref{fig2}.
Throughout this paper, we consider $T_{\rm L} = 600$~K ($k_{\rm B}T_{\rm L} \approx 0.05$~eV) and $T_{\rm R} = 300$~K ($k_{\rm B}T_{\rm R} \approx 0.025$~eV), a choice often adopted in the literature~\cite{Zebarjadi2012,Whitney2015,Hershfield2013}.

From the thermodynamic parameters $T_{\rm L}$, $T_{\rm R}$, $\mu_{\rm L}$, $\mu_{\rm R}$ we can construct the dimensionless parameter 
\begin{equation} \label{eq05}
   x \ \equiv \ \frac {\mu_{\rm R}-\mu_{\rm L} }{k_{\rm B}(T_{\rm L}-T_{\rm R})} \ ,
\end{equation} 
which is positive for $\mu_{\rm R}>\mu_{\rm L}$ and negative for $\mu_{\rm R}<\mu_{\rm L}$, having assumed $T_{\rm L}>T_{\rm R}$. 
It is easy to verify that 
\begin{equation} \label{eq06}   
   \frac{\varepsilon_0 - \mu_{\rm L}}{k_{\rm B}T_{\rm L}} 
   \ = \ \frac{\varepsilon_0 - \mu_{\rm R}}{k_{\rm B}T_{\rm R} } 
   \ = \ \frac{\mu_{\rm R} - \mu_{\rm L}}{k_{\rm B}(T_{\rm L} -T_{\rm R}) } 
 	 \ \equiv \  x \ ,
\end{equation}
thus $\varepsilon_0$ is at the right of both chemical potential for $\mu_{\rm R}>\mu_{\rm L}$ and at the left of both chemical potentials for $\mu_{\rm R}<\mu_{\rm L}$.
 
We now focus on thermoelectric transport through a device characterized by a narrow peaked transmission function at the resonance energy $E_d$. 
This situation typically occurs in the case of systems operating as energy filters~\cite{Humphrey2002} as quantum dots~\cite{Luo2016,Nakpathomkun2010}, quantum wells and quantum wires~\cite{Sothmann2013}.
For convenience, we describe such a resonance with a narrow rectangular-shaped transmission coefficient of the type 
\begin{equation} \label{eq07}
  {\mathcal T}(E) \ = \ \left\{ \begin{array}{rcl}
  & A_d \,\,\,\,\, & 
 {\rm if } \,\,\,\,\, E_d-\dfrac {\Gamma_d}{2} <E < E_d+\dfrac {\Gamma_d}{2}      
         \\[3mm]
    &0 \,\,\,\,\, & {\rm otherwise}\,.
     \end{array} \right. 
  \end{equation} 
We also assume that the transmission coefficient ${\mathcal T}(E)$ is rigid with respect to charge injection due to temperature and voltage gradients. Indeed, in realistic cases, in the presence of electron-electron and electron-phonon interactions ${\mathcal T}(E)$ should be determined self-consistently as a function~\cite{ArguelloLuengo2015} of $T_{\rm L}$, $T_{\rm R}$ and $V$. 
Nonetheless, this simple model is sufficient for our purpose, which is to show, without quantitative calculations, the existence of quantum bounds in the value of exchanged power and heat currents.

\begin{figure}[ht]
 \begin{center}
 \includegraphics[width=8cm]{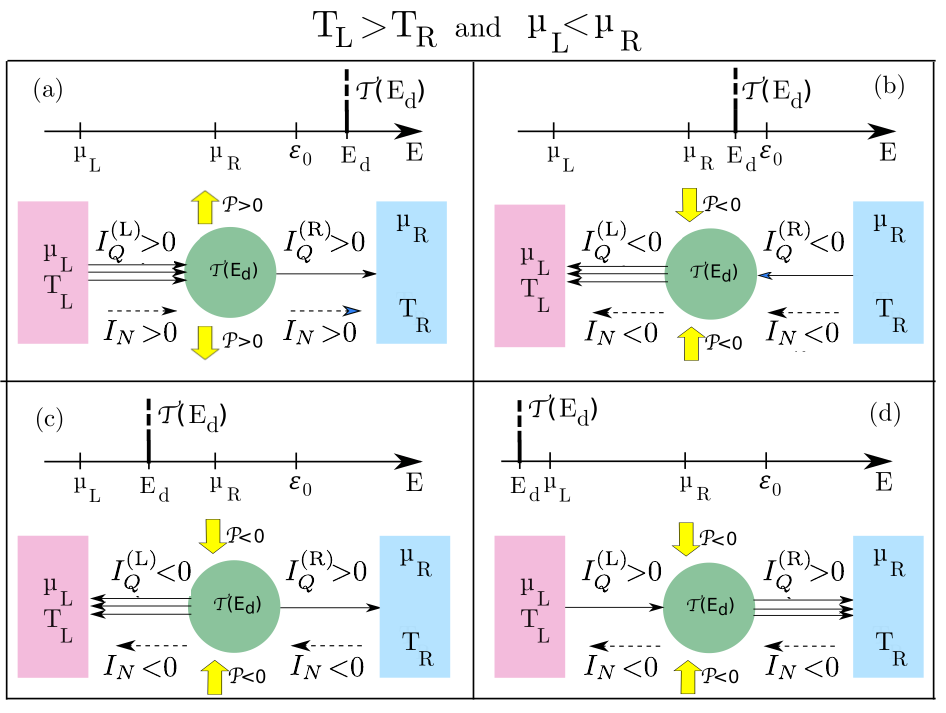}
 \end{center}
 \caption{ \label{fig3} Schematic representation of transport processes for the ideal filtering device in the configuration $T_{\rm L}>T_{\rm R}$ and $\mu_{\rm L} <\mu_{\rm R}$.
		 (a) Power generation regime for $\mu_{\rm L} < \mu_{\rm R} < \varepsilon_0 < E_d$. 
		 (b) Refrigeration regime for $\mu_{\rm L} < \mu_{\rm R} < E_d < \varepsilon_0 $. 
		 (c) Intermediate dissipative regime, for $E_d$ between the two chemical potentials. 
		 (d) Semi-infinite dissipative regime, for $E_d$ 
 smaller than both chemical potentials.} 
\end{figure}

By combining \eeref{eq01} and \eref{eq07}, and assuming $f_{\rm LR}(E) \approx f_{\rm LR}(E_d)$, which is reasonable for $\Gamma_d \ll k_{\rm B} T $, the particle, charge and heat currents become 
\begin{equation} \label{eq08}
   I_N \ = \ \frac{ A_d \ \Gamma_d }{h} \ f_{\rm LR}(E_d) \ ,
	 \ \ \ \ \ 
	 I_e \ = \ (-e)I_N\ ,
	\ \ \ \ \ 
	 I_{Q}^{\rm (L,R)} \ = \ \frac{(E_d - \mu_{\rm L,R})}{h} \ A_d \ \Gamma_d \ f_{\rm LR}(E_d) \ . 
\end{equation} 
For $E_d = \varepsilon_0$, the left and right thermal currents, the particle current and the power are all equal to zero. 
In fact, at the energy $\varepsilon_0$ the occupation states in the two reservoirs is the same and the Carnot efficiency is reached~\cite{Humphrey2002}.
We notice that \emph{left and right heat currents have different signs if the resonance $E_d$ lies in the interval between the chemical potentials, and the same sign otherwise}. 
Therefore, having $E_d$ between the two chemical potentials is not useful either for refrigeration or for power production.
According to \eref{eq01d} the power takes the expression 
\begin{equation}  \label{eq09}
   {\mathcal P} \ = \ \frac{1}{h} \, (\mu_{\rm R}- \mu_{\rm L}) \ A_d \ \Gamma_d \ f_{\rm LR}(E_d) \ . 
\end{equation} 
From \eref{eq09} we see that in the case $\mu_{\rm L} <\mu_{\rm R}$ and $f_{\rm LR}(E)$ corresponding to \fref{fig2}(a), the thermoelectric device generates energy (i.e., ${\mathcal P}>0)$ if $E_d>\varepsilon_0$, see \fref{fig3}(a), and absorbs energy (i.e., ${\mathcal P}<0)$ if $E_d<\varepsilon_0$, \fref{fig3}(b). 
Therefore, in the configuration ($T_{\rm L} > T_{\rm R}$; $\mu_{\rm L} <\mu_{\rm R}$) the device operates as power generator for $E_d>\varepsilon_0$ and as refrigerator for $E_d<\varepsilon_0 $.
\noindent In the case of the power generator, see \fref{fig3}(a), the thermal efficiency parameter for heat-to-power conversion becomes, 
\begin{equation} \label{eq10}
   \eta^{(tm)}(E_d) \ \equiv \ \frac{ \mathcal P }{ I_{Q}^{\rm (L)} } 
                    \    =   \ \frac{ \mu_{\rm R} -\mu_{\rm L} } { E_d -\mu_{\rm L} } \qquad {\rm with}
     \qquad \varepsilon_0 < E_d < \infty \ .
\end{equation} 
The maximum value of the efficiency parameter occurs for $E_d \equiv \varepsilon_0$ where it equals the efficiency of the Carnot cycle, in fact, from \eref{eq04} it follows $\eta^{(tm)}(E_d =\varepsilon_0)= (T_{\rm L}-T_{\rm R})/T_{\rm L}=\eta_c^{(tm)}$. 
At the same time, according to \eref{eq09}, the generated power ${\mathcal P} (E_d)$ vanishes.

In the case of the refrigeration regime, see \fref{fig3}(b), the coefficient of performance becomes
\begin{equation} \label{eq11}
 \eta^{(refr)}(E_d) \ \equiv \  \frac{ I_{Q}^{\rm (R)}(E_d) }{ \mathcal P}
                    \    =   \  \frac{E_d- \mu_{\rm R}} {\mu_{\rm R}- \mu_{\rm L}}
      \qquad {\rm for} \qquad \mu_{\rm R} \le E_d \le \varepsilon_0 \ .
\end{equation} 
The efficiency of the refrigeration machine is zero for $E_d=\mu_{\rm R}$, and takes the maximum value at $E_d=\varepsilon_0$ where it equals the coefficient of performance of the Carnot cycle for refrigeration: $\eta^{(refr)}(E_d =\varepsilon_0)=T_{\rm R}/(T_{\rm L} -T_{\rm R} )= \eta_c^{(refr)}$. 
     
When $E_d$ is smaller than one or both chemical potentials (\fref{fig3}(c) and \fref{fig3}(d)), power is absorbed and fully dissipated into heat transferred to both reservoirs with no useful result. 
Specifically, in the case $ \mu_{\rm L}<E_d<\mu_{\rm R}$, see \fref{fig3}(c), $f_{\rm LR}(E_d)<0 $, and we have: $I_Q^{\rm (L)} < 0, I_Q^{\rm (R)}> 0, {\mathcal P}< 0$ and $ I_N < 0$. The power is then absorbed and wasted into heat transferred to both reservoirs. We call this regime the \emph{intermediate dissipative regime}.
When the resonance energy $E_d$ is instead located in the energy interval [$-\infty, \mu_{\rm L} $] (\fref{fig3}(d), $f_{\rm LR}(E_d)<0 $ and we have: $I_Q^{\rm (L)} >0, I_Q^{\rm (R)} \gg 0, {\mathcal P} <0$ and $I_N <0$. 
The power is then absorbed, and wasted into heat transferred to the right reservoir. We call this regime the \emph{semi-infinite dissipative regime}.

\begin{figure}[ht]
 \begin{center}
 \includegraphics[width=8cm]{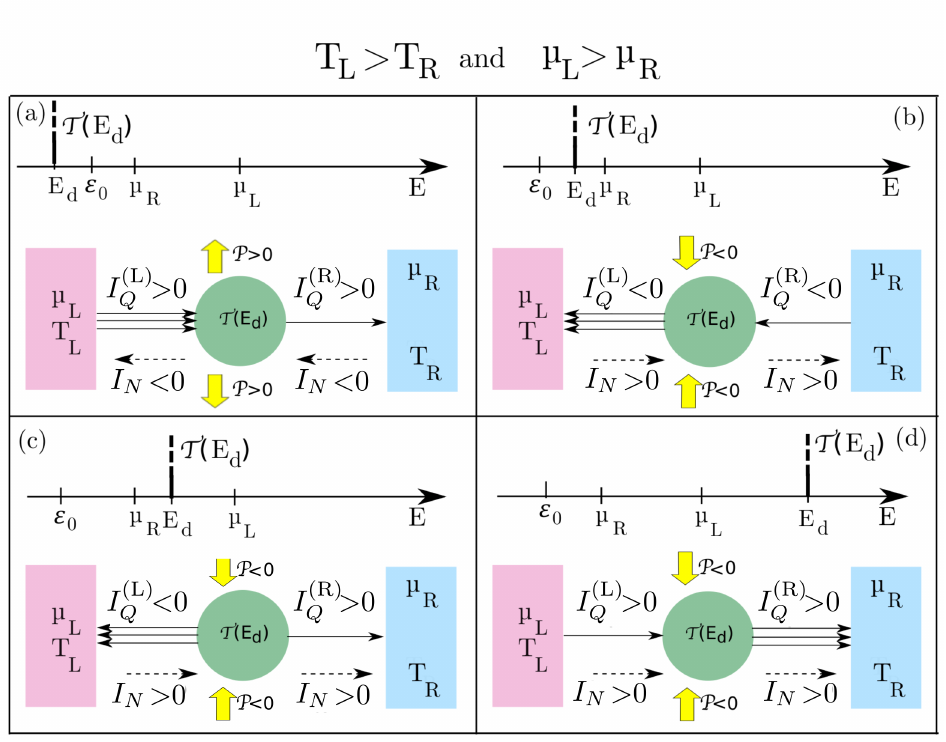}
 \end{center}
 \caption{ \label{fig4} Schematic representation of transport processes for the ideal filtering device in the configuration $T_{\rm L}>T_{\rm R}$ and $\mu_{\rm L} >\mu_{\rm R}$. 
		(a) Power generation regime for $E_d < \varepsilon_0 < \mu_{\rm R} < \mu_{\rm L}$. 
		(b) Refrigeration regime for $ \varepsilon_0 < E_d < \mu_{\rm R} < \mu_{\rm L}$. 
		(c) Intermediate dissipative regime, with $E_d$ between the two chemical potentials.
		(d) Semi-infinite dissipative regime, with $E_d$ larger than both chemical potentials.}
\end{figure}

Similar considerations hold in the case $T_{\rm L}>T_{\rm R} $ and $\mu_{\rm L} > \mu_{\rm R}$. The results are summarized in \fref{fig4}.  
 
\begin{figure}[b] 
 \begin{center}
 \includegraphics[width=8cm]{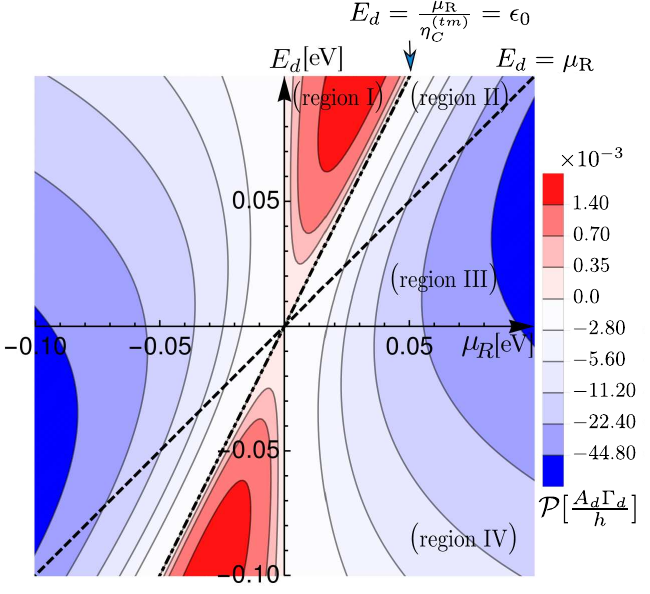}
 \end{center}
 \caption{\label{fig5} Contour plot in the $(E_d, \mu_{\rm R})$ plane of the power exchanged by the thermoelectric device with peaked transmission function. 
	The plot is invariant under inversion symmetry, and it is sufficient to focus on $\mu_{\rm R}>0$ and $E_d>0$. 
 The device works as power producing machine or refrigerator in regions~I and II, respectively. 
 Regions~III and IV represent useless dissipative regimes. }
\end{figure}

\section{Quantum bounds for the thermoelectric regimes of the ideal peaked-filtering device\label{sec3}}
The results of the previous section are pictorially summarized in \fref{fig5}, where the contour plot illustrates the power generated or absorbed, in units $A_d\Gamma_d/h$, as function of $E_d$ and $\mu_{\rm R}$.
We assume $\mu_{\rm L}$ as the reference energy and, without loss of generality, we set $\mu_{\rm L}=0$. 
Moreover, we indicate two particularly important lines in the $(E_d,\mu_{\rm R})$ plane. 
The bisector line $E_d = \mu_{\rm R}$ marks the border between the refrigeration regime (region~II) and the intermediate dissipative regime (region~III). 
The steeper line $E_d = \mu_{\rm R}/\eta_c = \varepsilon_0$ marks the border between the refrigeration (region~II) and the thermal (region~I) regimes. 
The $E_d$-axis marks the border between the power production and the semi-infinite dissipative regime (regions~II and IV), while the $\mu_{\rm R}$-axis marks the other border between the two dissipative regimes.
Thanks to the inversion symmetry of the exchanged power ${\mathcal P}$ with respect to the axis $(E_d,\mu_{\rm R})$, for the regions of interest I and II, we can restrict our considerations to the first quadrant in \fref{fig5}.
For $\mu_{\rm R}>0$ the power generation region~I is delimited by the constraints $ \mu_{\rm R} > 0$ and $E_d > \varepsilon_0={\mu_{\rm R}}/{\eta_c^{(tm)}}$, while the refrigeration region is delimited by the constraints $\mu_{\rm R} > 0$ and $\mu_{\rm R} < E_d < \varepsilon_0= {\mu_{\rm R}}/{\eta_c^{(tm)}}$.  
It is seen by inspection that \emph{in region~I the produced power ${\mathcal P}$ is bounded}, since 
\begin{equation} \label{eq12}
    {\mathcal P} \ = \ \frac{A_d \ \Gamma_d }{h} \, \mu_{\rm R} \ f_{\rm LR}(E_d) 
                 \ < \ \frac{A_d \ \Gamma_d }{h} \, \mu_{\rm R} \ f_{\rm L}(E_d) 
                 \ < \ \frac{A_d \ \Gamma_d }{h} \, \mu_{\rm R} \ f_{\rm L}\!\!\left( \frac{\mu_{\rm R}}{\eta_c^{(tm)}}\right) .
\end{equation} 
The last quantity in the above equation is evidently bounded as the chemical potential $\mu_{\rm R}$ is varied in the interval $[0,+\infty]$, and so is the power production. 
Exploiting the relation (\ref{eq05}) with $ \mu_{\rm L}=0$ one obtains the dependence of $ {\mathcal P}$ from the temperature difference:
\begin{equation} \label{eq13}
   {\mathcal P} \ < \ \frac{A_d \ \Gamma_d }{h} \ k_{\rm B} \ (T_{\rm L}-T_{\rm R}) \ \frac{x}{e^x+1}.
 \end{equation} 

The maximum of the above equation is found from the solution of a transcendent equation. 
The result is $ {\mathcal P} < \dfrac{A_d\Gamma_d }{h} k_{\rm B} \,(T_{\rm L}-T_{\rm R})\, 0.2785$. 
This occurs for $x$=1.2785, i.e., $ \mu_{\rm R}= 0.032$~eV. 

\begin{figure}[ht] 
 \begin{center}
 \includegraphics[width=15.5cm]{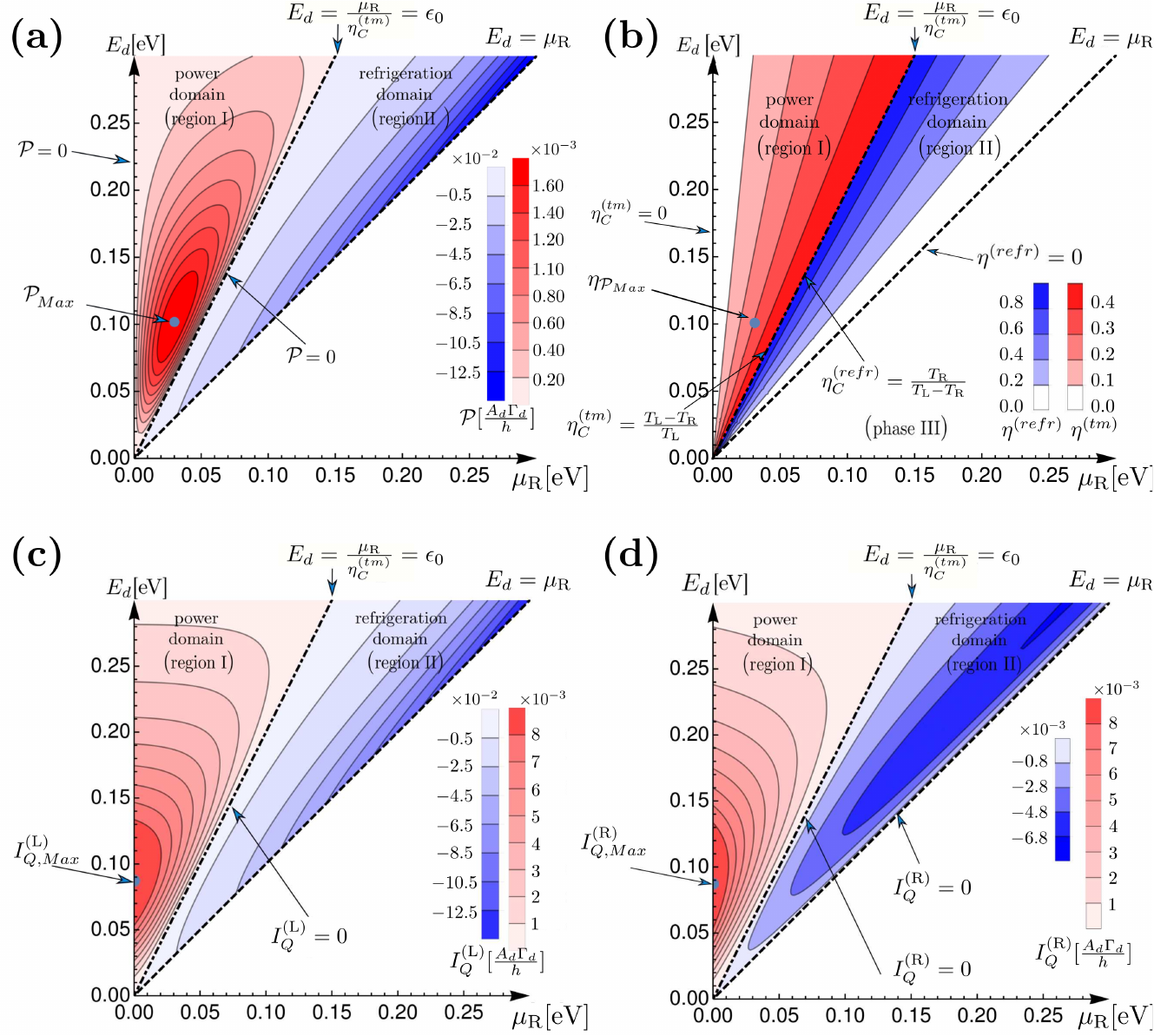}
 \end{center}
 \caption{\label{fig6} (a) Contour plot of the exchanged power by the thermodynamic device in regions~I and II. 
 The power production in region~I reaches the maximum value at $E_d =0.102$~eV and $\mu_{\rm R} = 0.030$~eV, and is zero along the $E_d$-axis and the $E_d= \varepsilon_0$ line. 
 The power absorbed in region~II is zero along the $E_d= \varepsilon_0$ line and becomes arbitrarily large along the bisector $E_d= \mu_{\rm R}$.
 (b) Plot of the efficiency and coefficient of performance in the two regions. 
			The efficiency at maximum produced power indicated in panel (a) is $\eta^{(tm)}_{\mathcal P_{Max}}$=0.296. 
 (c) Contour plot of the left thermal current $I_Q^{\rm (L)}$, which is positive and bounded in the thermal machine regime, and negative and unbounded in the refrigeration regime. 
			(d) Contour plot of the right thermal current $I_Q^{\rm (R)}$, which is positive in the power production region and negative in the refrigeration region. 
			The right thermal current is bounded in both regimes.}
\end{figure}

\Fref{fig6}(a) reports the contour plot of the power exchange for regions~I and II. 
For the chosen temperatures, the maximum generated output power occurs for $E_d =0.102$~eV and $\mu_{\rm R} = 0.030$~eV.    
As expected, ${\mathcal P}$ is zero along the $E_d$-axis, where $\mu_{\rm R}=\mu_{\rm L} (=0)$, and also along the line $E_d=\varepsilon_0 = \mu_{\rm R}/\eta_c^{(tm)}$, where $f_{\rm L} - f_{\rm R} = 0$.
Conversely, in the refrigeration region the absorbed power is negative and \emph{is not bounded}. 
In fact, the absorbed power along the $E_d = \mu_{\rm R}$ line (setting $\mu_{\rm L}=0$) reads 
\begin{equation} \label{eq14}
   {\mathcal P}(E_d=\mu_{\rm R}) \ = \ \frac{A_d \ \Gamma_d }{h} \, \mu_{\rm R} \ f_{\rm LR}(\mu_{\rm R})
                                 \ = \ \frac{A_d \ \Gamma_d }{h} \, \mu_{\rm R} \ \left[ \frac{1}{e^{\mu_{\rm R}/k_{\rm B}T_{\rm L}} + 1} - \frac{1}{2}\right] \ .
\end{equation} 
%
%
\noindent The above expression is evidently \emph{not bounded} for large values of the chemical potential $\mu_{\rm R}$.
In summary, ${\mathcal P}$ is zero along the line $E_d=\varepsilon_0 = \mu_{\rm R}/\eta_c^{(tm)}$, while it is given approximately by the value $(-1/2) (A_d\Gamma_d /h) \mu_{\rm R}$ along the line $E_d = \mu_{\rm R}$, when $\mu_{\rm R}$ exceeds few $k_{\rm B} T_{\rm L}$. 
   
Figure 6b reports the contour plot of the efficiency (red region) and the performance coefficient (blue region) of the considered ideal filtering nanostructure.
The line $E_d = {\mu_{\rm R}}/ {\eta_c^{(tm)}}$ provides the (maximum) Carnot efficiency both for heat-energy conversion and for refrigeration, which are $\eta_c^{(tm)}=0.5$ and $\eta_c^{(refr)}=1$ for the considered temperatures. 
However, the corresponding exchanged power is zero, see \fref{fig6}(a).
Around this line an optimal trade-off for power generation or refrigeration can be established. 
The efficiency at the maximum power output in region~I is $\eta^{(tm)}_{\mathcal P_{Max}} $=0.296. 

As for the heat current outgoing the hot left contact, we have $I_Q^{\rm (L)} = (A_d\Gamma_d /h) E_d \, [ f_{\rm L}(E_d) - f_{\rm R}(E_d) ]$.
Within region~I, $I_Q^{\rm (L)}$ is positive, and becomes zero along the border line $E_d= \varepsilon_0 = \mu_{\rm R}/\eta_c^{(tm))}.$ Most importantly, \emph{the left thermal current is bounded in the region~I} since in that region we have 
\begin{equation} \label{eq15}
     I_Q^{\rm (L)} \ < \ \frac{A_d \ \Gamma_d }{h} \ E_d \ f_{\rm L}(E_d) 
                   \ = \ \frac{A_d \ \Gamma_d }{h} \ E_d \ \frac{1}{e^{E_d/k_{\rm B} T_{\rm L}} +1} \ . 
\end{equation} 

\noindent From \fref{fig6}(c), it is seen that the maximum of the left thermal current occurs at the point $E_d \approx 0.091$~eV, along the border line $\mu_{\rm R}=0$.
This last feature is indeed expected: for $\mu_{\rm R}= \mu_{\rm L} (=0)$) no chemical potential barrier is of obstacle to the carrier diffusion.
 
In the refrigeration region~II, $I_Q^{\rm (L)}<0$ is not bounded.
Consider in fact \eref{eq08} for the left thermal current along the line $E_d=\mu_{\rm R}$. 
One obtains
\begin{equation} \label{eq16}
   I_Q^{\rm (L)}(E_d=\mu_{\rm R}) \ = \ \frac{A_d \ \Gamma_d }{h} \ \mu_{\rm R} \ \left[ f_{\rm L}(\mu_{\rm R}) - f_{\rm R}(\mu_{\rm R})\right]
	                                \ = \ \frac{A_d \ \Gamma_d }{h} \ \mu_{\rm R} \ \left[ \frac{1}{e^{\mu_{\rm R}/k_{\rm B}T_{\rm L}} + 1} - \frac{1}{2}\right] \ .
\end{equation} 
The last expression is evidently \emph{not bounded}, and so is the heat current flowing from the left reservoir. 
Along the line $E_d=\mu_{\rm R} \gg k_{\rm B}T_{\rm L}$, it holds $I_Q^{\rm (L)}\approx(-1/2)(A_d\Gamma_d /h)\mu_{\rm R}$.
As shown in \fref{fig6}(d), the coincidence of \eref{eq14} and \eref{eq16} is due to the fact that the right thermal current is zero on the boundary line $E_d=\mu_{\rm R}$. 
  
Eventually we consider the heat current flowing from/to the cold right reservoir, as reported in \fref{fig6}(d). 
In the power generation regime, the right thermal current $I_Q^{\rm (R)}$, see Eq.8, is $positive$ and $bounded$. 
In fact the quantity $(E_d -\mu_{\rm R})$ is always positive and then
\begin{equation} \label{eq17}
     I_Q^{\rm (R)} \ < \ \frac{A_d \ \Gamma_d}{h} \ (E_d -\mu_{\rm R}) \ f_{\rm L}(E_d) 
                   \ < \ \frac{A_d \ \Gamma_d}{h} \ E_d \ f_{\rm L}(E_d)
                   \ = \ \frac{A_d \ \Gamma_d}{h} \ E_d \ \frac{1} { e^{E_d/k_{\rm B}T_{\rm L}} + 1 } \ .
      \hspace{1cm} 
\end{equation} 

\noindent As mentioned above, $I_Q^{\rm (R)}$ in \fref{fig6}(d) and $I_Q^{\rm (L)}$ in \fref{fig6}(c) are perfectly equal on the $E_d$ axis since the power generated vanishes there. 
In the refrigeration region the right heat current $I_Q^{\rm (R)}$ is negative and bounded, in contrast to the left unbounded current.
In fact, from \eref{eq17} we obtain
\begin{equation} \label{eq18}
   I_Q^{\rm (R)} \ > \ - \frac{A_d \ \Gamma_d}{h} \ (E_d - \mu_{\rm R}) \ f_{\rm R}(E_d) 
             		 \ = \ - \frac{A_d \ \Gamma_d}{h} \ (E_d - \mu_{\rm R}) \ \frac{1} {e^{(E_d-\mu_{\rm R})/k_{\rm B}T_{\rm R}} + 1} \ .
\end{equation} 
Again $I_Q^{\rm (R)}$ vanishes along the lines $E_d=\varepsilon_0 = \mu_{\rm R}/\eta_c^{(tm)}$ and $E_d = \mu_{\rm R}$. 
 
\begin{table}[h]
 \caption{Quantum bounds for a nanoscale thermoelectric with peaked transmission function. \label{table1}
} 
\centering
{\begin{tabular*}{\textwidth}{@{\extracolsep{\fill}}cccc@{}}
\hline
\multicolumn{4}{c}{Power generation region \,\,\,\, $\mu_{\rm L}=0$ \,\,\,\, $\mu_{\rm R}>0$ \,\,\, \,$E_d>\varepsilon_0$} \\
\hline
\midrule
\midrule
\vspace{1mm}
${\mathcal P}>0$ & bounded & ${\mathcal P}<\dfrac{A_d\Gamma_d }{h} \, \mu_{\rm R} \, f_{\rm L}(E_d)$& \eref{eq12} \\ \vspace{1mm}
$ I_Q^{\rm (L)} >0$ & bounded &$ I_Q^{\rm (L)} < \dfrac{A_d\Gamma_d }{h} \, E_d \, f_{\rm L}(E_d)$ & \eref{eq15} \\ \vspace{1mm}
$I_Q^{\rm (R)} >0$ & bounded& $I_Q^{\rm (R)}< \dfrac{A_d\Gamma_d }{h} \, E_d \, f_{\rm L}(E_d)$& \eref{eq17} \\
\midrule
\hline
\multicolumn{4}{c}{Refrigeration region \,\,\, \,$\mu_{\rm L}=0$ \,\,\, \,$\mu_{\rm R}>0$ \,\,\, \,$\mu_{\rm R}<E_d<\varepsilon_0$} \\ 
\hline
\midrule
${\mathcal P}<0$ & not bounded & $|{\mathcal P}| \propto \mu_{\rm R} $& \eref{eq14}\\ \vspace{1mm}
$ I_Q^{\rm (L)} <0$ & not bounded & $|I_Q^{\rm (L)}(E_d=\mu_{\rm R})| \propto \mu_{\rm R} $& \eref{eq16} \\ \vspace{1mm}
$I_Q^{\rm (R)} <0$ & bounded & $|I_Q^{\rm (R)}|<\dfrac{A_d\Gamma_d }{h} \, (E_d -\mu_{\rm R})\, f_{\rm R}(E_d)$& \eref{eq18}\\
\hline
\bottomrule
\end{tabular*}} {}
\end{table}
 

\section{Conclusions\label{sec4}}
 
We investigated the thermoelectric transport properties of a nanoscale system characterized by a transmission coefficient peaked at the energy $E_d$, connected to two particle reservoirs at different temperatures and chemical potentials. 
We identified the regions in the parameter space set $T_{\rm L}$ , $T_{\rm R}$, $\mu_{\rm L}$, $\mu_{\rm R}$ and $E_d$ where the system works as a thermal machine or as a refrigerator with the corresponding efficiency and coefficient of performance, and provided a simple demonstration of the existence of quantum bounds for the exchanged power and heat currents.
In particular, as summarized in \tref{table1}, we have shown that the power produced by the thermal machine is bounded and positive, and that the thermal currents $I_Q^{\rm (L)}$ and $I_Q^{\rm (R)}$ are both positive and bounded, with a single maximum of the same value located at the same position along the $E_d$-axis. 
In contrast, for the refrigerator the absorbed power is negative and not bounded. 
We also notice the absence of bounds in $I_Q^{\rm (L)}$. On the contrary $I_Q^{\rm (R)}$ is strictly bounded, in fact heat current out of a reservoir at temperature $T$ cannot exceed appropriate quantum bounds~\cite{Pendry1983}. 
Analytic results in the literature exploiting more realistic shapes of the transmission function~\cite{Whitney2014,Whitney2015,Rey2007,Luo2018} are in qualitative agreement with the results here obtained with a simple and intuitive procedure for the peaked transmission model.
 

\begin{acknowledgments}
G.~M.~acknowledges the financial support from EDISON-Volta Prize (2018).
\end{acknowledgments}
  

\bibliography{manuscript}

\end{document}